# Ultrafast solid-state chemical synthesis of BaTiO$_3$ initiated by gyrotron microwave radiation

S.V. Sintsov*, N.V. Chekmarev, K.I. Rybakov, A.A. Sorokin, E.I. Preobrazhenskii, A.V. Vodopyanov

*sins@ipfran.ru

A.V. Gaponov-Grekhov Institute of Applied Physics of the Russian Academy of Sciences, Nizhny Novgorod 603950, Russia

**Abstract**

This work presents the results of a study on the solid-state synthesis of barium titanate under continuous microwave radiation from a 24 GHz gyrotron in a multimode cavity reactor. It is shown that in localized domains where fine-scale thermal instabilities develop, initiated by microwave radiation within the initial stoichiometric reaction mixture of ultrafine barium carbonate and titanium dioxide powders, the synthesis can proceed within 1,5 - 7 minutes, achieving a target product yield of up to 90%. Based on a developed realistic numerical model of the multimode reactor, involving an iterative solution of stationary Maxwell and heat conduction equations, it is demonstrated that the specific absorbed power in the domains where fine-scale thermal instabilities develop can reach 670 W/cm³ under an input microwave power of 400 W.

**Introduction**

High-temperature solid-state chemical synthesis is one of the key processes for producing a wide range of functional inorganic materials [1]. Methods based on conventional heating of reaction mixtures are characterized by significant thermal inertia, high energy consumption, and prolonged synthesis times, often amounting to many hours. These factors not only increase production costs but can also lead to a reduction in product quality due to intensive grain growth and the formation of impurity phases during prolonged high-temperature exposure [2, 3].

The duration of solid-state physicochemical processes can be substantially reduced by switching to microwave heating [4-6]. Its primary advantage lies in the volumetric heating mechanism, where the electromagnetic field energy is directly absorbed by the material and converted into heat. This approach can result in achieving extremely high heating rates, reducing the overall energy consumption significantly compared to conventional methods, and in some cases allowing the formation of target phases at lower temperatures. It also holds potential for obtaining products with enhanced properties, such as smaller grain size or higher phase purity [4-14]. However, the practical implementation and industrial scaling of these advantages for solid-state processes in common microwave systems, employing magnetrons at a radiation frequency of 2.45 GHz, face significant challenges. A fundamental limitation stems from the non-uniform distribution of the electromagnetic field within powdered media. This, along with variations in the electrodynamic and thermodynamic parameters of the reaction mixture depending on temperature, can lead to uncontrolled local overheating ("hot spots"), compromising synthesis uniformity and hindering both process control and scalability.

This work explores a novel approach to microwave solid-state synthesis, based on the purposeful initiation and use of fine-scale thermal instabilities. Its essence lies in employing shorter-wavelength (millimeter-wave) electromagnetic radiation to create highly contrasting temperature fields within the reaction mixture – specifically, strongly localized domains of extreme overheating. Crucially, conditions can be achieved where these domains are thermally isolated from the bulk material due to the inherently low thermal conductivity of the medium, attainable in reaction mixtures of loose powders. The sharp temperature gradient arising at the boundaries of these fine-scale thermal instabilities enables spatial separation of the final product (within the high-temperature domains) and the unreacted initial material. This facilitates high reaction rates and selectivity without overheating the entire volume.

In this work, this mechanism is experimentally demonstrated for the solid-state reaction synthesis of barium titanate ($BaTiO_3$). The reaction was initiated by heating an initial stoichiometric mixture of dispersed barium carbonate ($BaCO_3$) and titanium dioxide ($TiO_2$) powders with continuous gyrotron microwave radiation at 24 GHz. To investigate the spatial temperature distribution within the medium exhibiting fine-scale thermal instabilities, a realistic numerical model of the experimental multimode microwave reactor has been developed.

**Experimental setup**

Figure 1 shows the schematic and photograph of the multimode microwave reactor. The reactor is a closed electrodynamic structure connected to the circular waveguide output transmission line of a 24 GHz gyrotron with a power of up to 5 kW in continuous-wave mode [15-17]. The microwave input window consists of a circular quartz plate, integrated into the waveguide transmission line. This window serves to protect the radiation source in the event of a gas discharge developing within the reactor and propagating towards the microwave radiation source.

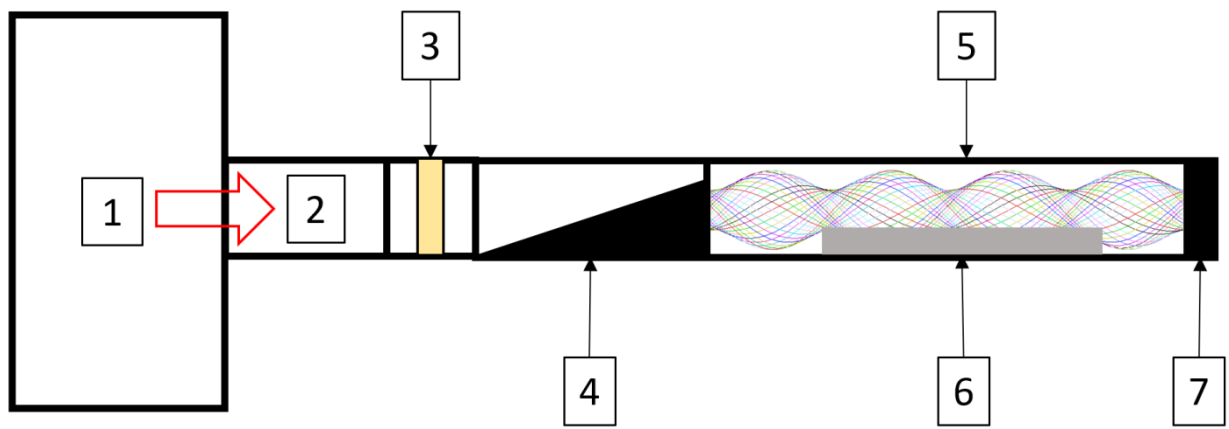
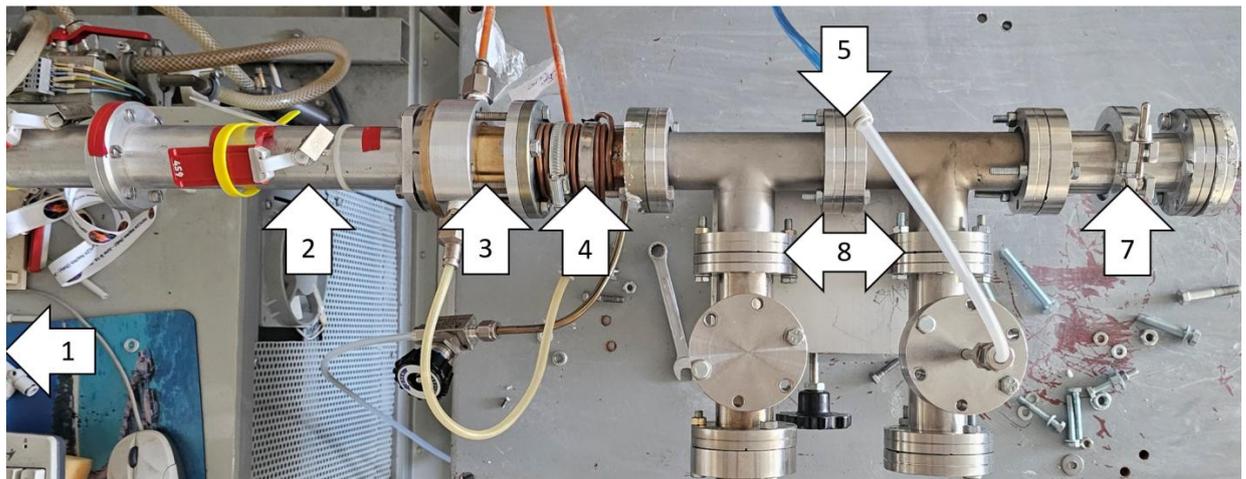

Figure 1. Schematic and photograph of the multimode microwave reactor. 1 – 24 GHz gyrotron; 2 – Circular waveguide transmission line (diameter 32.6 mm); 3 – Quartz input window (quarter-wave thickness); 4 – Microwave filter; 5 – Multimode microwave reactor chamber; 6 – Reaction powder mixture on an I-beam shaped aluminum substrate; 7 – Quick-release flange port for powder loading/unloading; 8 – Functional flange ports (auxiliary connections).

A critical element of the setup, electrodynamically bounding the reactor volume, is the microwave filter. This filter comprises a tapered section of a circular waveguide and is designed to prevent microwave radiation from propagating backwards to the source. Additional reduction in reflected power is achieved by orienting the filter's output slit perpendicular to the polarization plane of the propagating mode.

The reactor chamber is constructed from two vacuum cross fittings with standard flange ports. In the described configuration, the unused ports were utilized for installing purge gas inlet/outlet assemblies and for the visual observation of the heating process. Temperature measurements were performed using a two-color infrared pyrometer through a viewport equipped with a wire mesh screen. Loading and unloading of the reaction mixture was accomplished via a quick-release flanged port. The reaction medium was a stoichiometric mixture of dispersed powders: barium carbonate (CAS No: 513-77-9) and titanium dioxide (CAS No: 13463-67-7). The mixture was prepared by mixing the components using pestle and mortar for 10 minutes. The powder mixture, with no compaction, was placed on an aluminum I-beam shaped substrate (24 × 50 mm) as a thin layer (3–4 mm thick). After weighing, it was positioned within the reactor volume for subsequent microwave heating.

**Description of the reaction medium model**

The process of solid-state chemical synthesis under microwave heating can be described by a mathematical model coupling self-consistent electrodynamic and thermal problems based on Maxwell's equations and the heat conduction equation, respectively. The electrodynamic problem calculates the distribution of the electromagnetic field within the reactor volume. The resulting field intensity distribution within the heated sample is then used to compute the volumetric heat source power density for the heat conduction equation.

A critical aspect for performing these calculations is the accurate assignment of the electrodynamic and thermophysical properties of the substances involved in the synthesis reaction. The materials used, which remain disperse powder media throughout all stages of the process, can be described as multiphase systems containing one or several different solid phases and a gaseous phase corresponding to the void space between particles. When the characteristic scale of the system's microstructural inhomogeneity is much smaller than the electromagnetic wavelength within the material, methods based on introducing averaged effective properties are employed to describe the interaction of the electromagnetic field with such materials.

Since the volumetric phase fractions in the processes considered here can vary widely, the effective properties of the materials are conveniently described using the effective medium approximation [18]. This approximation assumes that spherical inclusions of each phase are embedded within a homogeneous medium possessing the sought-after effective properties. For the complex permittivity of the powder medium, this approach leads to the following equation:

$$\sum_i C_i \frac{\varepsilon_i - \varepsilon_{\text{eff}}}{2\varepsilon_{\text{eff}} + \varepsilon_i} = 0, \qquad (1)$$

where $\varepsilon_i$ is the complex permittivity of the i-th component of the mixture (including voids), and $C_i$ is its relative volume fraction. This algebraic equation, whose degree equals the number of different mixture components, can be solved for the sought effective complex permittivity $\varepsilon_{\text{eff}}$. A similar equation defines the effective thermal conductivity $\lambda_{\text{eff}}$ of the powder medium.

The starting materials for the synthesis process in this work were an equimolar mixture of loose powders: $BaCO_3$ + $TiO_2$. The modeling assumed a two-stage process: first, the decomposition of $BaCO_3$ occurs at 800 K:

$$BaCO_3 \rightarrow BaO + CO_2\uparrow, \qquad (2)$$

and then the synthesis of $BaTiO_3$ proceeds at 1073 K:

$$BaO + TiO_2 \rightarrow BaTiO_3. \qquad (3)$$

Therefore, the initial system is three-phase ($BaCO_3$, $TiO_2$, and air in the voids). The intermediate system, for which calculations are performed in the temperature range 800–1073 K, is also three-phase (BaO, $TiO_2$, and air). The final system is two-phase ($BaTiO_3$ and air). Data on the volume fractions of the mixture components were calculated based on measurements of the initial powder density. Data on the volume fraction (relative density) of the densifying $BaTiO_3$ were taken from experiments on its microwave sintering at various temperatures [19].

Data on temperature-dependent high-frequency dielectric properties of materials are often scarce in the literature. For the calculations described herein, data for the permittivity of $BaCO_3$ were taken from [20], for the real part of the high-frequency permittivity of $TiO_2$ from [21], and for the high-frequency dielectric loss of $TiO_2$ from [22]. Data for the high-frequency permittivity and dielectric loss of BaO were taken from [23]. For $BaTiO_3$, temperature-dependent data on high-frequency permittivity and dielectric loss from [24] were used; these data were obtained during microwave heating experiments on loose powder and they account for the density changes due to sintering. Extrapolation of available dependencies to higher frequencies was used where necessary.

Instead of thermal conductivity data for $BaCO_3$, temperature-dependent data for $CaCO_3$ from [25] were used. Potential distortions of the results due to this substitution are insignificant, as they can only influence the calculated temperature field during the initial heating stage. Data on the temperature-dependent thermal conductivities of BaO and $TiO_2$ were taken from [21]. For $BaTiO_3$, data from [26] were used, extrapolated to the high-temperature region while accounting for changes in relative density during sintering. Temperature-dependent data for the thermal conductivity of air were taken from [24].

The specific heat capacity of the mixtures was calculated using an additive method, considering the molar ratios between the mixture components. Data on the temperature-dependent specific heat capacities of the components were taken from [21] and [27].

The resulting temperature dependencies of the effective parameters of the powder media used in the modeling are shown in Figure 2.

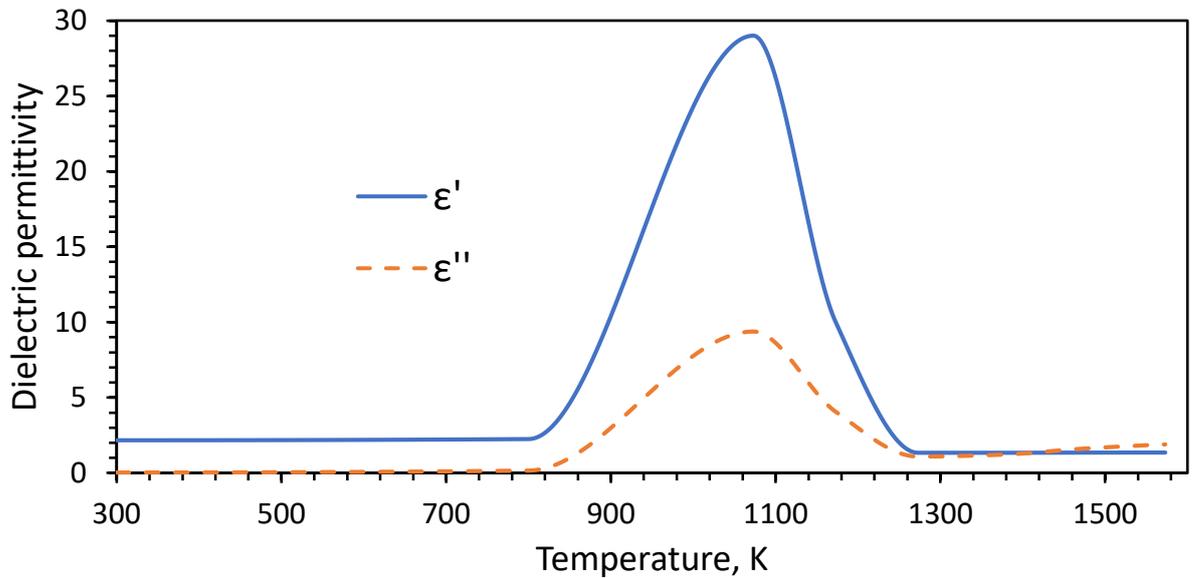

(a)

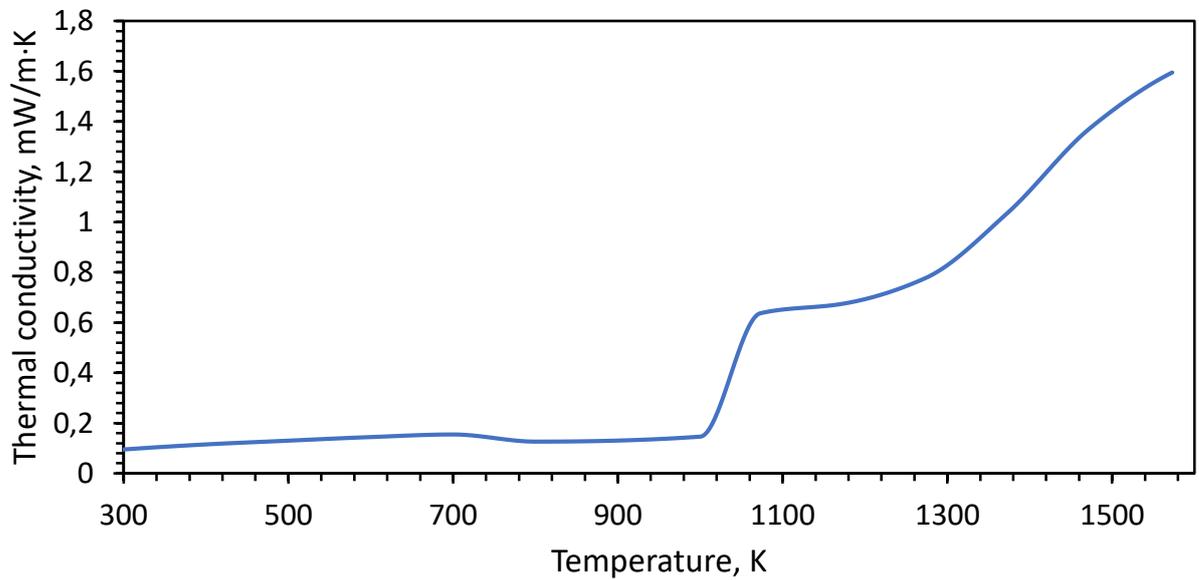

(b)

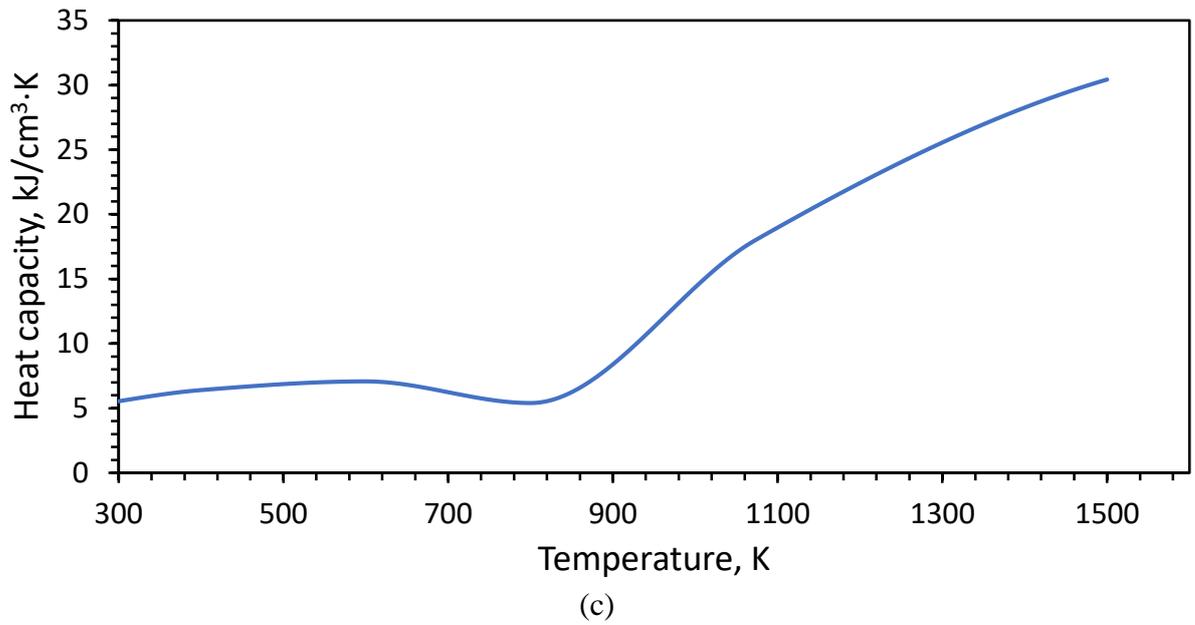

(c)

Figure 2. Temperature dependencies of the effective parameters of the powder media used in modeling: a) Complex permittivity; b) Thermal conductivity; c) Heat capacity per unit volume.

**Numerical modeling**

To investigate the mechanisms of fine-scale thermal instability development during the heating of a powdered reaction mixture by microwave electromagnetic radiation, a realistic three-dimensional finite element model of the experimental microwave multimode reactor (Figure 3) was developed. In addition to the main structural elements, the model includes a circular waveguide port on the right boundary, excited by the $TE_{11}$ mode at 24 GHz with linear polarization.

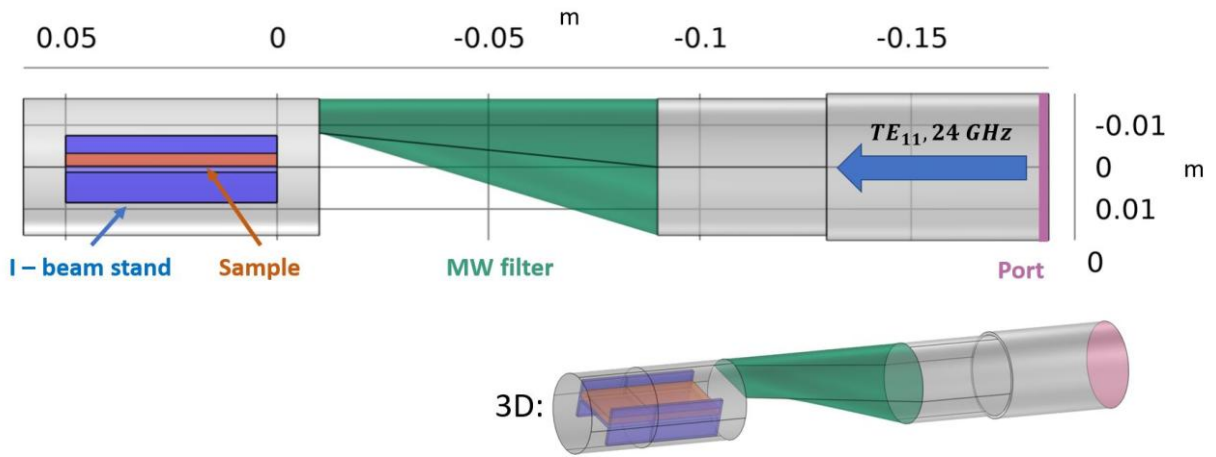

Figure 3. Realistic three-dimensional finite element model of the multimode microwave reactor.

The self-consistent steady state problem, combining the calculation of electromagnetic wave propagation and absorption with heat transfer modeling, was solved using an iterative procedure. This procedure computes the volumetric heat source power density and the resulting temperature distribution within the medium. An impedance boundary condition, defined by Eq. (4), was applied to the external boundaries of the computational domain and the surface of the metallic substrate:

$$[\boldsymbol{n} \times (\boldsymbol{E} - Z_s \cdot [\boldsymbol{n} \times \boldsymbol{H}])] = 0, \quad (4)$$

where $Z_s$ is the surface impedance of the material, **n** is the unit normal vector to the material surface, and **E** and **H** are the intensities of the electric and magnetic components, respectively, of the microwave field. This condition adequately accounts for the energy absorption within the thin skin layer of metallic elements, not requiring its explicit resolution in the mesh. The radiation reflected from the reactor is absorbed at the input port boundary, where its modal composition is analyzed to determine the reflection coefficients (S-parameters).

Heat transfer is modeled within the volume of the powder sample jointly with the aluminum I-beam shaped substrate. Heat sources include volumetric Joule losses, calculated from the distribution of absorbed microwave power in the dielectric sample, and losses in the metallic reactor walls due to the skin effect. Combined radiative heat transfer and free convection conditions to the surrounding environment, modeling heat removal from the system, are applied to the external boundaries of the sample and substrate.

The simulation results obtained for the case of an input power of 400 W demonstrate the following key features. The microwave electric field intensity distribution (Figure 4) exhibits a quasi-periodic standing wave structure above the sample. The positions of the electric field intensity maxima directly correspond to the domains of maximum microwave energy absorption and, consequently, the most intense heat generation. The calculated power balance shows a heating efficiency of above 75 %: out of 400 W of input power, 303 W is absorbed within the powder sample itself, 3 W is lost in the walls, and 94 W is reflected back.

An analysis of the absorbed power density (Figure 5a) and temperature (Figure 5b) distributions reveals their strong spatial correlation. A number of localized domains in which the temperature is sufficient for the solid-state reactions to occur (T > 1300 K) form on the sample surface. These high-temperature domains, are arranged quasi-periodically with a characteristic distance of about

7 mm between adjacent maxima. This distance is close to half the free-space wavelength of the radiation (λ = 12.5 mm).

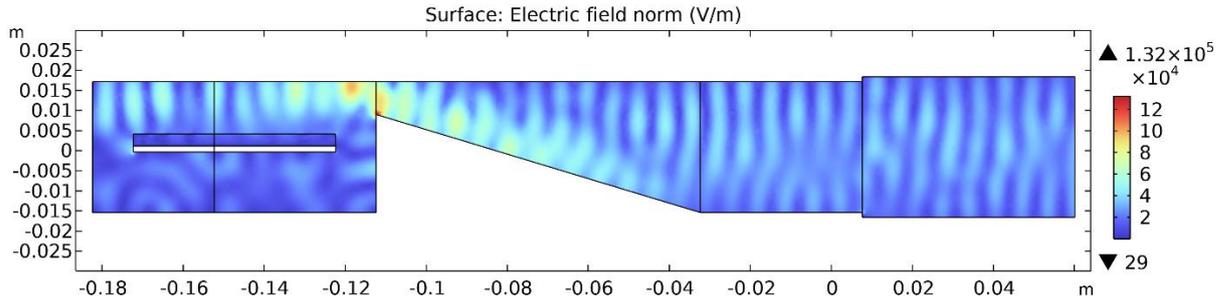

Figure 4. Electric field intensity distribution in the multimode microwave reactor model.

Depending on the magnitude of the specific power absorption, the characteristic size of the high-temperature domains varies between 3 and 8 mm. Changes in the sample position do not qualitatively alter the electric field distribution pattern within the reactor volume; rather, the thermal instability domains shift relative to the sample.

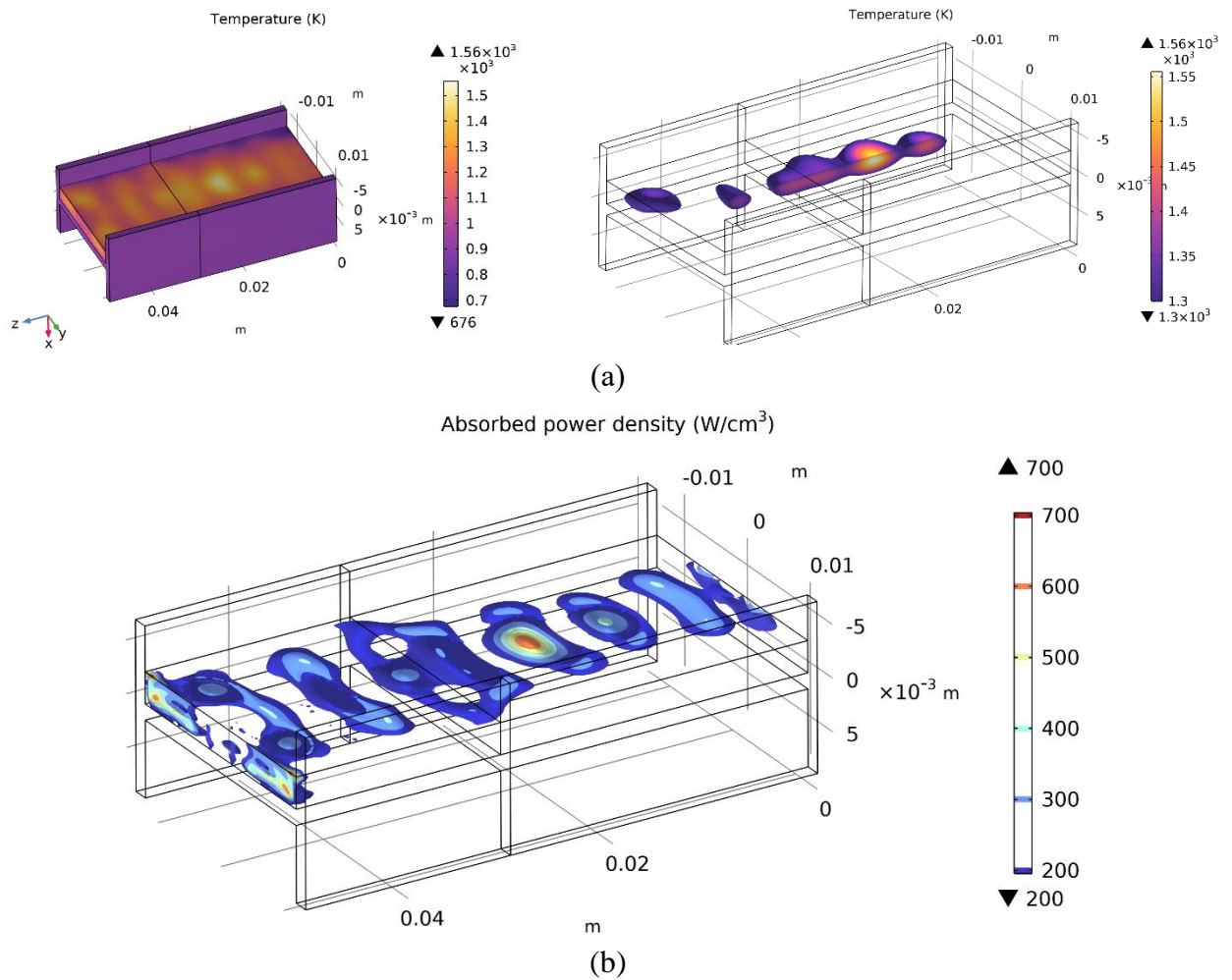

Figure 5. Temperature distribution (a) and absorbed power density distribution (b) in the sample placed within the multimode microwave reactor. The input power is 400 W.

The calculated specific absorbed power reaches 670 W/cm³. A characteristic value of the temperature gradients at the boundaries of the high-temperature domains is 60 K/mm. As the input power increases, the maximum specific power absorption within the domains of thermal instability increases nonlinearly (Figure 6). Concurrently, the characteristic size of the high-temperature domains also increases, eventually reaching a state where neighboring domains merge, forming a continuous layer of the target reaction product. For all power values, the temperature within the simulation space exceeds 1300 K, signifying the transition beyond the temperature-dependent permittivity function adopted for the model (Figure 2a).

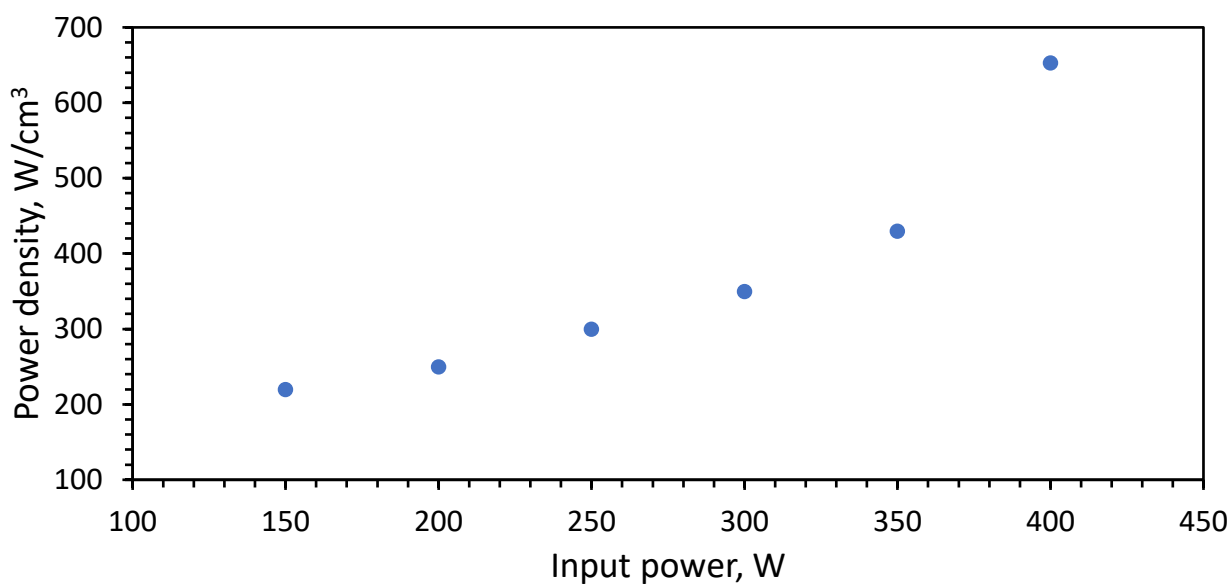

Figure 6. Maximum volumetric microwave power density within thermal instability domains vs. the input microwave power.

**Experimental results**

In the experiments, the solid-state synthesis process was initiated by applying microwave radiation with the input power varied between 200 W and 400 W. After 30–80 s, red-glowing spots appeared on the surface of the powder layer, indicating the development of thermal instabilities in these domains. The process duration was 1,5–7 minutes, after which the microwave source was switched off. In the areas where red spots appeared, the surface temperature measured 600–800 °C, depending on the input power, and remained constant during further heating.

As a result of the process, agglomerated structures formed beneath the surface of the powder layer. These structures, measuring approximately 3–6 mm in size, were located precisely in the regions where red spots had appeared during heating. Increasing the power and heating time led to the merging of these structures and the formation of larger agglomerates. Figure 7 presents characteristic photographs of the powder layer and the agglomerates extracted from it after heating in the multimode microwave reactor. The mass loss determined by weighing the samples before and after processing amounted to 13–15%. This loss is attributed to the release of carbon dioxide during the decomposition of barium carbonate and confirms indirectly that the achieved temperature has been sufficient for the solid-state synthesis of barium titanate.

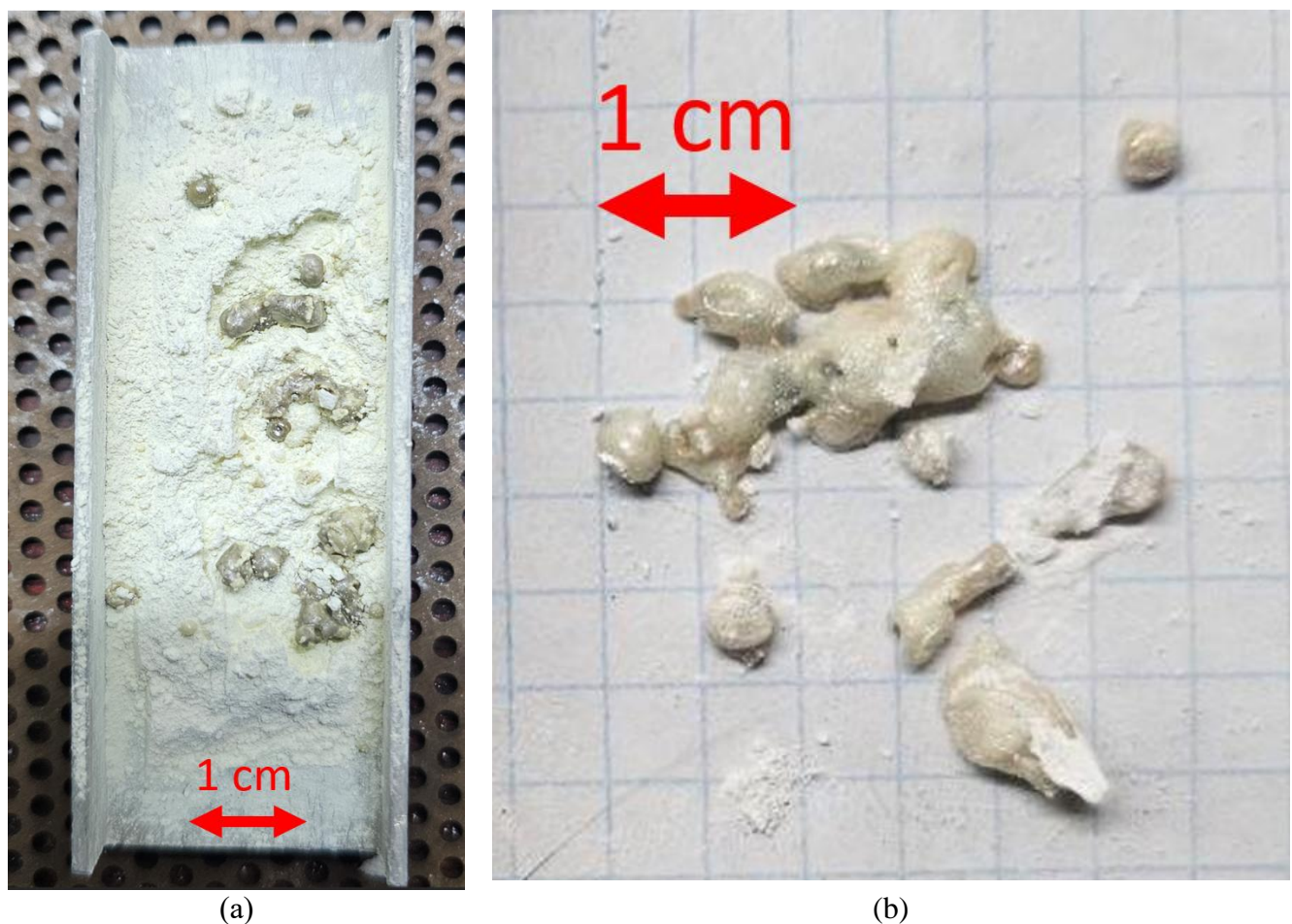

(a)                                  (b)

Figure 7. Photographs of the powder layer on the I-beam shaped aluminum substrate (a) and the agglomerates extracted from it (b) after heating in the multimode microwave reactor. Input microwave power: 400 W, heating time: 7 minutes.

The obtained agglomerates were mechanically crushed for X-ray diffraction (XRD) analysis performed on an Ultima IV diffractometer (Rigaku, Japan) using CuKα radiation. Phase identification was carried out using the quantitative analysis based on corundum numbers. Figure 8 shows a characteristic XRD pattern of the crushed agglomerates synthesized in the multimode microwave reactor.

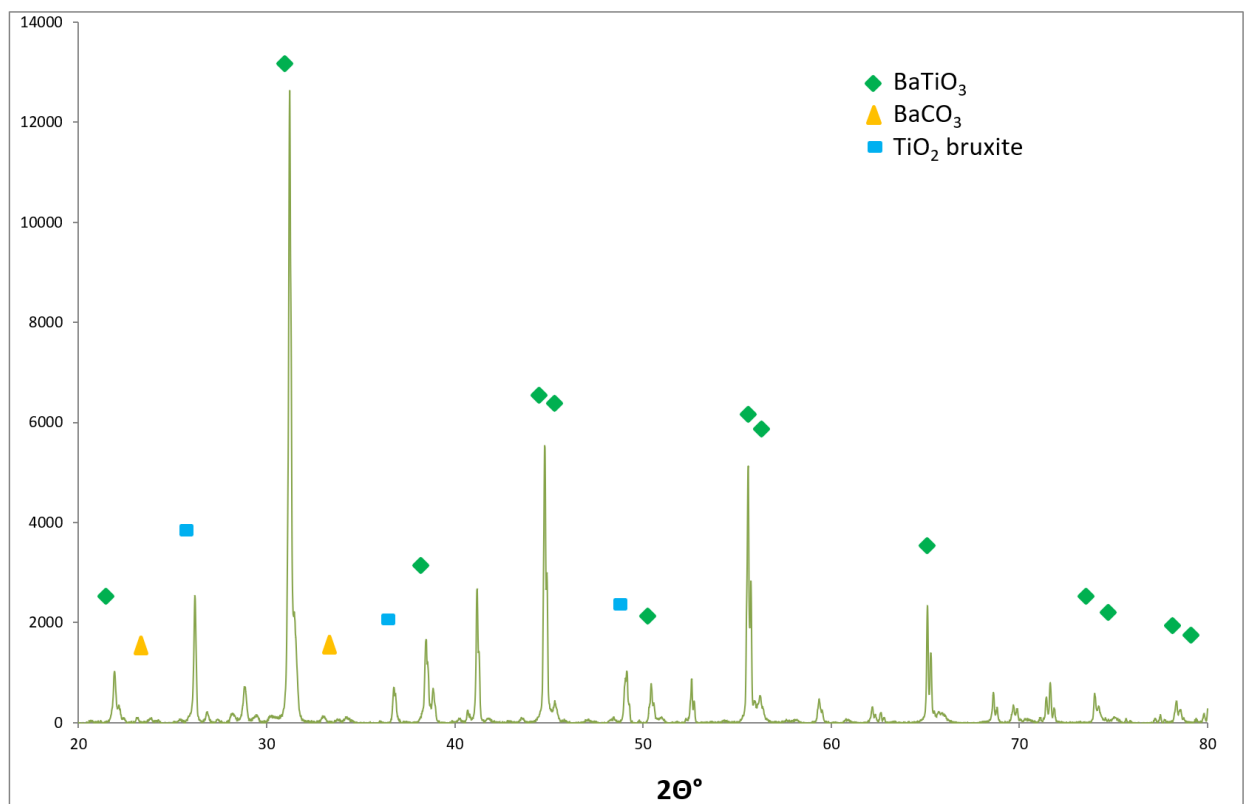

Figure 8. Characteristic XRD patterns of crushed agglomerates synthesized in the multimode microwave reactor. The synthesis regime: 400 W, 7 minutes.

Detailed analysis revealed a significant dependence of the phase composition on the processing regimes. At microwave power levels of 200-300 W and synthesis times of 1.5-4 minutes, a complex multi-component composition formed, dominated by various polymorphic modifications of $BaTiO_3$. The presence of the monoclinic phase $BaO(TiO_2)_2$ in amounts of 6-7% appears to indicate local barium deficiency, caused by kinetic diffusion limitations under the conditions of ultrafast synthesis. Increasing the power to 300-400 W and the processing time to 5-7 minutes led to a substantial change in the phase composition. Under these conditions, the tetragonal modification of $BaTiO_3$ dominated (up to 68-72%), indicating a more complete reaction progression and an approach to a thermodynamically equilibrium state. The presence of the hexagonal phase (18-23%), stable at temperatures above 1430°C, suggests local overheating of the reaction zones to temperatures significantly exceeding the average sample temperature. At maximum heating power, the formation of a small amount (up to 3% for a process duration of 5 minutes) of non-stoichiometric compounds such as $Ba_2TiO_4$ and $Ba_2Ti_9O_{20}$ was also observed, characterized by a deficiency of barium and oxygen relative to titanium. The residual amounts of barium carbonate (up to 5-7%) observed in some cases may be related to the specifics of decarbonization kinetics under ultrafast heating conditions. Microstructural analysis showed the formation of dense sintered agglomerates with clear boundaries corresponding to the zones of maximum heat release. The absence of signs of nanoscale coherent scattering regions or amorphous phases in all samples indicates complete crystallization of the synthesis products even under ultrafast heating conditions. Of particular interest is the spatial correspondence between the experimentally observed agglomerate sizes (3-6 mm) and the calculated sizes of the overheating instability zones obtained by numerical modeling. This correspondence not only confirms the accuracy of the developed numerical model but also indicates the deterministic nature of the overheating instability formation process.

**Discussion**

The obtained experimental and calculated data demonstrate a fundamental difference of the proposed approach from traditional methods of microwave synthesis. While most research in the field of microwave material processing is aimed at suppressing temperature field inhomogeneities [4, 6, 12], this work proposes a paradigm of purposeful creation and use of controlled fine-scale thermal (overheating) instabilities. This became possible due to the use of millimeter-wave radiation, which ensures strong spatial and energy localization of the overheating zones.

A key advantage of the method is the achievement of extremely high heating rates in localized domains while maintaining the bulk of the material at significantly lower temperatures. This creates conditions where the kinetics of the solid-state reaction are determined not by classical diffusion-controlled equations (Jander, Ginstling-Brounshtein) but by processes initiated under conditions of sharp temperature gradients (up to 60 K/mm). The obtained values of specific absorbed power (up to 670 W/cm³) are an order of magnitude higher than typical values for traditional microwave systems at 2.45 GHz [6, 12], where the typical power density does not exceed 50-100 W/cm³ even in "hot spots". This enables a reduction of synthesis time to just a few minutes, whereas in studies on microwave synthesis of barium titanate at 2.45 GHz, it is almost an order of magnitude longer [6].

XRD phase analysis showed the formation of highly crystalline barium titanate with a main phase content of 88-96%, which is comparable to the best literature data for traditional synthesis methods [4]. However, unlike convective methods leading to the formation of large crystalline aggregates [2, 3] and conventional microwave heating often causing non-uniform grain growth [12], the described method ensures the formation of localized agglomerates 3-6 mm in size with a homogeneous particle distribution. Such a microstructure is explained by the short duration of thermal exposure and the presence of sharp temperature gradients, hindering particle coalescence processes.

It is necessary to note that the developed approach has certain limitations related to the thermal stability of the synthesized materials. Exceeding a power of 700 W leads to the melting of barium titanate (T > 1500 K), which requires further refinement of scalable models of the described method. One scalable approach could be the use of quasi-optical beam transformation systems for millimeter-wave radiation, allowing for the joint variation of microwave power density and the size of the reaction zone as it moves along the beam focal line [13, 15].

The prospects of the developed approach are associated with the possibility of controlling not only the kinetics of chemical reactions but also the microstructure of the obtained materials by varying the radiation parameters. Further development of the method may include the use of controlled microwave phase shifters to create specified spatial distributions of temperature fields, which would allow realizing the principles of additive manufacturing in the microwave synthesis of functional materials.

**Conclusion**

A novel approach to microwave solid-state chemical synthesis has been developed, based on the intentional initiation and use of fine-scale thermal (overheating) instabilities. A high degree of conversion of the initial reagents into the target product (over 90%) within domains with developed overheating instability has been demonstrated experimentally. Based on a realistic numerical model of the multimode microwave reactor, characteristic values of the specific absorbed power in the localized instability domains and the temperature gradient achieved at their boundaries have been obtained. It has been shown that varying the parameters of microwave heating allows for the targeted control of the phase composition of the resulting product. Under optimal processing regimes, the predominant formation of the tetragonal modification of $BaTiO_3$ (up to 70-75%) is achieved, representing the greatest practical interest for ferroelectric applications. Microstructural analysis has confirmed the formation of finely dispersed agglomerates with a homogeneous particle distribution. This is due to the suppression of coalescence processes, which is caused by the short duration of thermal exposure and the presence of sharp temperature gradients at the boundary of the reaction zone. Realizing the potential of the developed method for creating energy-efficient and high-speed technologies of solid-state chemical synthesis requires further study of the reaction kinetics under conditions of extremely rapid local heating and optimization of microwave reactor systems.

**Acknowledgements**

The research on solid-state chemical synthesis under microwave heating was financially supported by the Foundation for Assistance to Small Innovative Enterprises in Science and Technology (FASIE), Contract No. 5242ГС1/101612 dated November 07, 2024.

The modeling of the effective properties of densifying reaction media under microwave heating was supported by the Russian Science Foundation (RSF), Grant No. 23-19-00363.